\documentclass
[floatfix,showpacs,amsmath,amssymb,aip,jmp,reprint]{revtex4-1}

\usepackage{amsmath}
\usepackage{amssymb}
\usepackage{overpic}
\usepackage{dsfont}
\usepackage{rotating}
\usepackage{xcolor}
\usepackage[urlcolor=blue,colorlinks=true,linkcolor=blue]{hyperref}

\newcommand{\ket}[1]{\ensuremath{|#1\rangle}}

\newcommand{\smal}{\scriptscriptstyle}

\begin{document}

\title{Riccati equation and the problem of decoherence}
\date{27 January  2010}

\author{Bart{\l}omiej Gardas}
\email{bartek.gardas@gmail.com}

\affiliation{Institute of Theoretical and Applied Informatics, Polish Academy
of Sciences, Ba{\l}tycka 5, 44-100 Gliwice, Poland}
\begin{abstract}

The block operator matrix theory is used to investigate the problem of a single qubit.  
 We will establish a connection between the Riccati operator equation and the possibility of obtaining an exact reduced dynamics for the qubit in question. The model of the half spin particle in the rotating magnetic field coupling with the external environment is discussed.
 We show  that the model defined in such a way can be reduced to a time independent problem. 
\end{abstract}
\pacs{03.65.Yz, 03.67.-a}   
\maketitle


\section{Introduction} \label{sec:intro}
Exactly solvable models for decoherence play an important role both in the theory of the open quantum system and quantum information theory~\cite{assym,qudit}. Unfortunately, most of the models describing the process of decoherence can not be solved exactly. However, there is a wide class of models for which exact reduced dynamics~\cite{alicki2} is known. Those models deal with the case where the energy transfer between the system and the environment is not present. This phenomenon is known as pure decoherence or dephasing~\cite{dephasing}. It has been  found that generalization of the dephasing models to the case where energy is exchanged between the system and the environment is straightforward, but for most of this generalization analytical solutions were not obtained. It is only natural to wonder why the dephasing models can be solved easily whereas even the most basic generalizations pose such a difficult task for the scientists.

In this manuscript we  show that obtaining the exact reduced dynamics of the model of one qubit interacting with the environment is at least as difficult as solving the Riccati operator equation associated with the Hamiltonian defining the model. First we will discuss the procedure allowing one to obtain the density matrix for the system using the block operator matrix perspective. 

The general form of the Hamiltonian describing the qubit $Q$  coupling with the external environment(heat bath) $E$  can be written as follows~\cite{alicki2} 
\begin{equation}
  H_{QE}=H_{Q}\otimes 1_{E}+1_{Q}\otimes H_{E}+H_{int},
\end{equation}
where $H_{Q}$, $H_{E}$ are the Hamiltionian of the qubit and the environment respectively and $H_{int}$ represents the interaction between the systems. Hamiltonian $H_{QE}$ acts on  $\mathcal{H}_{Q}\otimes\mathcal{H}_{E}$ space, where $\mathcal{H}_{Q}$ and $\mathcal{H}_{E}$ are the Hilbert spaces for the system and the environment respectively. For most models it is assumed that the initial state of the $Q+E$ system has the following form $\rho_{QE}=\rho_{Q}\otimes\rho_{E}$, which means that there is no correlation between $Q$ and $E$ initially (see~\cite{korelacje} and ref. therein). Our analysis is free of this assumption. The state of the $Q$ system at any given time $t$ takes the form:
\begin{equation}
  \rho_{Q}(t) =\mbox{Tr}_{E}(U_t\rho_{Q}\otimes\rho_{E} U_t^{\dagger}), \label{r:trace}
\end{equation}
where $U_t$ is the evolution operator of the $Q+E$ system and by $\mbox{Tr}_{E}(\cdot)$ we denote the partial trace. The $\rho_{Q}(t)$ is called reduced dynamics (with respect to the degree of freedom of the environment). From now on the quantity $\rho_{Q}(t)$ will be called the solution of the model. In the case of  $\mathcal{H}_{Q}=\mathbb{C}^2$ and $\mathcal{H}_{E}=\mathcal{H}$, where $\mathcal{H}$ is the arbitrarily separable Hilbert space (in general $\mbox{dim}\mathcal{H}=\infty$) the following isomorphism holds $\mathbb{C}^2\otimes\mathcal{H}=\mathcal{H}\oplus \mathcal{H}$. Therefore, any given operator $A$ acting on the  $\mathbb{C}^2\otimes\mathcal{H}$ space can be thought of as the $2\times 2$ block operator matrix (BOM) $[A_{ij}]$, where $A_{ij}, (i,j=1,2)$ act on  $\mathcal{H}$. In this notation the procedure of calculating partial trace $\mbox{Tr}_{E}$ is very intuitive, namely
\begin{equation}\label{r:partial}
\mbox{Tr}_{E}(A)=
\left[ 
\begin{array}{cc}
\mbox{Tr} A_{11} & \mbox{Tr}A_{12}   \\
 \mbox{Tr}A_{21} & \mbox{Tr}A_{22} 
\end{array}
\right],
\end{equation}
where $\mbox{Tr}(\cdot)$ is a trace on $\mathcal{H}$.
One can easily see that obtaining reduced dynamics  $\rho_{Q}(t)$ is very simple. However, the equation~(\ref{r:trace}) is far less useful than its theoretical simplicity might indicate. The reason is that one can not determine the exact block operator $2\times 2$ matrix form of the evolution operator $U_t$ of the system $Q+E$. The task becomes even more difficult when the Hamiltonian is time dependent.

There have been few theories resolving the problem of finding reduced dynamics both for time dependent and time independent Hamiltonians~\cite{alicki2,davis}. Furthermore a majority of the scientists focus their effort on a numerical methods and on perfecting the approximation methods~\cite{nonmarkov}. As a consequence  most of the research on the quantum information theory is based on a numerical rather then an analytical approach.   As a result during past several years no progress has been made in solving the known models.

The main purpose of this manuscript is to present an analytical approach. We consider one of the most established and useful models, namely the spin $1/2$ (qubit) in the rotating magnetic field. In the case where no coupling with the external environment is present, an analytical solution can be found in an elegant and simple manner~\cite{thaller}. If the mentioned coupling (modelled by quantum system of infinite number of degree of freedom) is present; however, the exact solution has not been found yet. We will not address this in the current manuscript; however, we will show that this model can be effectively reduced to the time independent problem (Section~\ref{sec:spin}). Moreover, we will show that the solution of any given model with a time independent Hamiltonian requires solving the Riccati operator equation associated with the Hamiltonian $H$ defining the problem (Section~\ref{sec:ric}). In other words, we will establish the connection between the problem of decoherence in physics and the mathematical problem of resolving the Riccati operator equation.  Furthermore, using the results of Section~\ref{sec:ric} we will discuss the possibility of obtaining an exact solution to the analyzed  problem from the set of differential equations on $\mathcal{H}\oplus\mathcal{H}$ (Section~~\ref{sec:diff}). Finally, in Section~\ref{sec:example} we consider a spin-bozon model as an example. Section~\ref{sec:summary} is a summary of the paper.


\section{Spin half in a rotating magnetic field and in contact with environment } \label{sec:spin}
Let us consider a single qubit in rotating magnetic field interacting with its environment. The qubit-environment time-dependent Hamiltonian reads
\begin{equation}
H(t,\beta) =H_{Q}(t,\beta)\otimes1_{E}+I_2\otimes H_{E}+H_{int},  \label{r:hamiltonian} 
\end{equation}
where $H_{Q}(t,\beta)$ and $H_{E}$ are Hamiltonians of qubit $Q$ and the environment respectively and $H_{int}$ represents the interaction between $Q$ and the  environment. It is assumed that $H_{int}$ takes the form $f(\sigma_3)\otimes V$, where $V$ is a Hermitian operator acting on $\mathcal{H}_{E}$ and $f(\sigma_3)$ is an analytic function of $\sigma_3$.
Hamiltonian $H_{Q}(t,\beta)$ is given by
\begin{equation}
H_{Q}(t,\beta) = \beta\sigma_3+\alpha\left(\sigma_1\cos\left(\omega t\right)+\sigma_2\sin\left(\omega t\right)\right), \label{r:time}
\end{equation} 
and it represents a spin system in rotating magnetic field $\vec{B}(t)$, where
\begin{equation}
\vec{B}(t) = \left[B_1\cos\left(\omega t\right), B_1\sin\left(\omega t\right),B_0\right].
\end{equation}
Here, $\alpha =\tfrac{1}{2}\omega_1\sim B_1$ and $\beta = \tfrac{1}{2}\omega_0\sim B_0$, where $B_0$, $B_1$ are amplitudes of the magnetic field~\cite{galindo2}.

The model described by the Hamiltonian~(\ref{r:hamiltonian}) cannot be solved exactly in this general case. By this we mean that the exact reduced dynamics $\rho_{Q}(t)$ for that model are not known. Let us now focus on another model defined by the Hamiltonian $H(\beta)\equiv H(0,\beta)$, where $H(t,\beta)$ is given by~(\ref{r:hamiltonian}). 

We will show that if  $\eta_t$ is a solution to the model described by the Hamiltonian~(\ref{r:hamiltonian}), and $\rho_t(\beta)$ represents a solution of the model with Hamiltonian $H(\beta)$, then the following equation holds
\begin{equation}
\eta_t = V_t\rho_t\left(\beta-\frac{\omega}{2}\right)V_t^{\dagger}, \label{r:rho}
\end{equation}
where
\begin{equation}
V_t = \mbox{diag}\left(e^{-i\omega t/2},e^{i\omega t/2}\right). \label{r:v}
\end{equation}
From equation~(\ref{r:rho}) and ~(\ref{r:v}) we see that if reduced dynamics $\rho_t(\beta)$  is known then all one needs to do to obtain the solution to the model of the $H(t,\beta)$ Hamiltonian is to introduce an effective parameter $\beta_{eff}:=\beta -\tfrac{\omega}{2}$, replace $\beta$ by $\beta_{eff}$, and perform a unitary transformation~(\ref{r:v}).
Since the procedure explained above is very simple we can effectively reduce the problem of solving model~(\ref{r:hamiltonian}) to one of solving the model $H(\beta)$.

In order to prove the equation~(\ref{r:rho}) let us note that the Hamiltonian~(\ref{r:hamiltonian}) satisfies the following condition $(\hbar=1)$
\begin{equation}
H(t,\beta) = e^{iKt}H(\beta)e^{-iKt}, \label{r:cond}
\end{equation}
where $K= -\tfrac{\omega}{2}\sigma_3\otimes 1_{E}$. This can be easily proven using the Baker-Campbell-Hausdorff formula~\cite{galindo}.
As was shown, in~\cite{thaller} every quantum system with Hamiltonian $H(t,\beta)$ satisfying~(\ref{r:cond}) for some Hermitian operator $K$ there evolves  
\begin{equation}
 U_t(\beta) = e^{iKt}e^{-iH_{eff}(\beta)t},\quad H_{eff}(\beta): = H(\beta)+K. \label{r:u}
\end{equation} 
Note that in general $[H(\beta),K]\not=0$ and therefore $[H_{eff}(\beta),K]\not=0$. In our case, from equation~(\ref{r:hamiltonian}) we learn that $H(\beta)=\left(\beta\sigma_3+\alpha\sigma_1\right)\otimes 1_{E}$, thus
\begin{eqnarray}
 H_{eff}(\beta) &=& \left(\beta\sigma_3+\alpha\sigma_1\right)\otimes 1_{E} -\frac{\omega}{2}\sigma_3\otimes 1_{E} \\ \nonumber
            &=& \left(\left(\beta-\frac{\omega}{2}\right)\sigma_3+\alpha\sigma_1\right)\otimes 1_{E} \\ \nonumber
            &=& H(\beta-\frac{\omega}{2}). \label{r:efektywny}
\end{eqnarray}
From equations~(\ref{r:u}) and~(\ref{r:efektywny}) we have
\begin{equation}
 U_t(\beta)= e^{iKt}U_t(\beta-\tfrac{\omega}{2}), \label{r:evolv}
\end{equation}
where $U_t(\beta)$ is the evolution operator generated by $H(t,\beta)$.
  Let $\hat{\rho}_t(\beta)$ and $\hat{\eta}_t$ be a density operator for the closed system  $Q+E$ associated with Hamiltonian $H(\beta)$ and $H(t,\beta)$ respectively in arbitrary time $t$. Let us also assume that $\hat{\rho}_0(\beta)=\hat{\eta}_0\equiv \hat{\rho}$. Using equation~(\ref{r:evolv}) one can easily see that
  
  \begin{eqnarray}\label{hat} 
   \hat{\eta}_t & = & U_t(\beta)\hat{\rho}U_t^{\dagger}(\beta) \\ \nonumber
                & = & e^{iKt}U_t(\beta-\tfrac{\omega}{2})\hat{\rho} U_t^{\dagger}(\beta-\tfrac{\omega}{2})e^{-iKt} \\ \nonumber
                & = & \hat{V}_t\hat{\rho}_t(\beta-\tfrac{\omega}{2})\hat{V}_t^{\dagger},    
  \end{eqnarray}
where we introduced $\hat{V}_t=e^{iKt}$.
To end the proof we will show that if $\hat{A}_1$, $\hat{A}_2\in B(\mathcal{H}\oplus\mathcal{H})$ are a $2\times2$ block operator matrix of the form $\hat{A}_i=A_i\otimes 1_{E},(i=1,2)$ and $\hat{B}=[\hat{B}_{ij}]\in B(\mathcal{H}\oplus\mathcal{H})$ then

\begin{equation}
 \mbox{Tr}_{E}(\hat{A}_1\hat{B}\hat{A}_2)= A_1\mbox{Tr}_{E}(\hat{B})A_2.  \label{tr}
\end{equation}
Equation~(\ref{tr}) follows from the linearity of trace $\mbox{Tr}$ operation and definition~(\ref{r:partial}) of partial trace. Note that $\hat{V}_t=V_t\otimes 1_{E}$, where $V_t$ is given by equation~(\ref{r:v}), thus taking partial trace of equation~(\ref{hat}) and using~(\ref{tr}) we obtain~(\ref{r:rho}) with $V_t$ given by~(\ref{r:v}).


\section{Operator Riccati equation}\label{sec:ric}
So far we have shown that the solution $\eta_t$ can be easily constructed from  $\rho_t(\beta)$. Now, we will pay attention to the possibility of obtaining an exact solution $\rho_t(\beta)$. 
 Let us now rewrite Hamiltonian $H(\beta)$ as a  block operator matrix~\cite{bom} 
 \begin{equation}\label{r:beta}
  H(\beta) = 
  \left[
\begin{array}{cc}
 H_{\smal{+}}+\beta & \alpha   \\
 \alpha & H_{\smal{-}}-\beta  
\end{array}
\right], 
 \end{equation}
 where we introduced $H_{\smal{\pm}}=H_{E}\pm V$.
Since Hamiltonian~(\ref{r:beta}) is time-independent, we can write the evolution operator as $ U_t = \mbox{exp}\left(-iH(\beta)t\right)$. We see that the main problem here is how to write down $U_t$ as $2\times 2$ BOM. 

If $\alpha =0$ this problem is trivial. On the other hand for $\alpha\not=0$ the diagonalization  of $2\times 2$ BOM is required  which is not a trivial problem~\cite{ricc}. 
With every Hermitian $2\times2 $ BOM of the form:
\begin{equation}
  R = 
  \left[
\begin{array}{cc}
 A & B   \\
 B^{\dagger} & C  
\end{array}
\right],  \quad A, B, C\in\mathcal{H}
\end{equation}
we can associate the \emph{operator} \emph{Riccati} \emph{equation}~\cite{riccati}
\begin{equation}
 XBX+XA-CX-B^{\dagger}=0, \label{r:riccati}
\end{equation}
where $X\in\mathcal{H}$. Solution $X$ of the equation~(\ref{r:riccati}), if it exists  can be used to construct $2\times2$ BOM: 
\begin{equation}\label{r:Ux}
U_{X}=
\left[
\begin{array}{cc}
 1_{E} & -X^{\dagger} \\
 X & 1_{E}
\end{array}
\right],
\end{equation} 
in such a way
\begin{equation}\label{r:invers}
   U_{X}^{-1}RU_{X}= 
     \left[
\begin{array}{cc}
 A+BX & 0   \\
 0 & C-B^{\dagger} X^{\dagger} 
\end{array}
\right].
\end{equation}
From the above consideration we see that to diagonalize  Hamiltonian~(\ref{r:beta}) we have to solve the following Riccati equation:
\begin{equation}
 \alpha X^2+X(H_{\smal{+}}+\beta)-(H_{\smal{-}}-\beta)X-\alpha = 0. \label{r:riccati2}
\end{equation}
Unfortunately, we do not know how to do that. Note that if $\alpha=0$ then $X=0$ is a solution. This is obvious since in that case $H(\beta)$ is already in the diagonal form. Note also that even if $\beta=0$ this problem is still very complicated.
 
 \section{Differential equation approach } \label{sec:diff}

Let us now transform the problem of solving a Riccati equation~(\ref{r:riccati2}) to the problem of solving a Schr$\ddot{\mbox{o}}$dinger equation on $\mathcal{H}\oplus\mathcal{H}$, with the Hamiltonian given by~(\ref{r:beta}). 
Let $\ket{\Psi_t}=[\ket{\psi_t},\ket{\phi_t}]^t$, then $\ket{\Psi_t}$ satisfy $i|\dot{\Psi}_t\rangle=H(\beta)\ket{\Psi_t}$. Of course, we can always
 write $\ket{\Psi_t}=\exp(-iH(\beta)t)\ket{\Psi_0}$, but this form of the solution is useless since $U_t$ does not have a $2\times 2$ BOM form. 
It may seem that we circled back to the point where we started since in writing the state $\ket{\Psi_t}$ as a column vector we need to diagonalize the matrix $H(\beta)$. Nothing could be further from the truth. To see this let us introduce operators $U$ and $J_t$ in the following way:
\begin{equation}
U =\frac{1}{\sqrt{2}}
\left[
\begin{array}{cc}
 1 & 1   \\
 1 & -1 
\end{array}
\right], 
\quad J_t=\exp (i\alpha\sigma_3t).
\end{equation}
 Let us also define $|\tilde{\Psi}_t\rangle =J_tU\ket{\Psi_t}$, and we can easily see that $i|\dot{\tilde{\Psi}}_t\rangle =H_t|\tilde{\Psi}_t\rangle $, where the periodic Hamiltonian is given by
\begin{equation}
H_t=
\left[
\begin{array}{cc}
 H_{E} & z_t^{\ast}(V+\beta)   \\
 z_t(V+\beta) & H_{E} 
\end{array}
\right], \quad z_t=e^{-i2\alpha t}.
\end{equation} 
The Riccati equation associated with $H_t$ reads
\begin{equation}
X(z_t^{\ast}V_{\beta})X+XH_{E}-H_EX-z_tV_{\beta}=0, \label{r:Triccati}
\end{equation} 
where $V_{\beta}=V+\beta$. Straightforward calculations show that $X_t=z_t$ is a solution of the Riccati equation~(\ref{r:Triccati}). 
According to~(\ref{r:invers}) we have
 
 \begin{equation}\label{r:sbom}
 S_t^{\dagger}H_tS_t=
 \left[
\begin{array}{cc}
 H_{+}+\beta & 0   \\
 0 & H_{\smal{-}}-\beta 
\end{array}
\right],
\end{equation}
where $S_t=\tfrac{1}{\sqrt{2}}U_{z_t}$ and $U_{z_t}$ is given by~(\ref{r:Ux}), namely:

\begin{equation}
S_t=\frac{1}{\sqrt{2}}
\left[
\begin{array}{cc}
 1 & -z_t^{\ast}   \\
 z_t & 1 
\end{array}
\right].
 \end{equation} 
Note that $S_t$ is a unitary $2\times 2$  block operator matrix.
We see that if one could solve the Schr$\ddot{\mbox{o}}$dinger equation for $|\tilde{\Psi}_t\rangle$ then our problem would be solved. Formally, we can always do that using chronological operator T, the solution is given by~\cite{chrono}
\begin{equation}
 |\tilde{\Psi}_t\rangle=\mbox{T} \exp (-i\int_0^tH_{\tau}\,d\tau) |\tilde{\Psi}_0\rangle . \label{r:sol}
\end{equation}
Sadly, the form~(\ref{r:sol}) of the solution has little use due to the presence of the chronological operator T. Nevertheless, it is interesting to note that the connection between the models~(\ref{r:beta}) and~(\ref{r:sbom}) is well define and the solution to the equation~(\ref{r:Triccati}) can be easily found, yet finding the solution to the equation~(\ref{r:riccati2}) poses a big problem.

\section{Examples} \label{sec:example}
Up until now we did not choose the specific form of the operators $H_E$ and $V$, which means that the analysis we presented was very general. That fact implies an important concept, namely that the analysis that we carried out does not depend on the particular choice of a heat bath. It is crucial, however that the coupling of the qubit with the environment is given by the matrix $f(\sigma_3)$, where $f$ is an arbitrary analytical function. It is interesting to consider the model where operators $H_E$ and $V=V(g)$ are defined as follows:
\begin{equation}
H_{E} = \int\limits_0^\infty\, d\omega\,\omega\, a^{\dagger}(\omega)a(\omega), \label{r:hr}
\end{equation}
where $a^{\dagger}(\omega)$ and $a(\omega)$ are bozonic annihilation and creation operators respectively and they satisfy the commutation relations: $\left[a(\omega), a^{\dagger}(\omega ')\right]=\delta (\omega -\omega ')$,\, $\omega$, $\omega '>0$.
$V(g)$ is given by
\begin{equation}
V(g)=\int\limits_0^\infty\, d\omega\,\left(g^{\ast}(\omega)a(\omega)+g(\omega)a^{\dagger}(\omega)\right), \label{r:vg}
\end{equation}
where $g\in L^2[0,\infty]$. Operators $H_E$ and $V(g)$  given by~(\ref{r:hr}) and~(\ref{r:vg}) define the bozonic heat bath~\cite{alicki2} of the qubit. One can find that
\begin{eqnarray}
H_{\smal{+}} & = & W(g)H_EW(g)^{\dagger}+C(g), \\
H_{\smal{-}} & = & W(g)^{\dagger}H_EW(g)+C(g), 
\end{eqnarray}
where $C(g)$ is a certain constant. One can always rescale the Hamiltonian so that $C(g)=0$, thus we will omit the constant. The unitary Weyl's operator  has the form $W(g)=\mbox{exp}(A(g))$, where
\begin{equation}
 A(g) = \int\limits_0^\infty\, d\omega\,\left(g^{\ast}(\omega)a(\omega)-g(\omega)a^{\dagger}(\omega)\right).
\end{equation}
In the case of the $\alpha=0$ model can be solved exactly~\cite{dephasing}. If $\alpha\not=0$ obtaining the exact reduced dynamics, according to~(\ref{r:riccati2}) is at least as difficult as solving the equation (to simplify we put $\beta=0$)
\begin{equation}
 \alpha X^2+X(WH_{E}W^{\dagger})-(W^{\dagger}H_{E}W)X-\alpha=0. \label{r:wriccati}
\end{equation}
 The solution of the equation ~(\ref{r:wriccati}) is yet to be discovered.
 
As a second examples let us consider a pure decoherence case. In this situation $[H_{Q}\otimes1_{E},H_{int}]=0$. Let $H_{int}=M\otimes V$, where $M$ is a arbitrary Hermitian $2\times 2$ matrix. Since operators $H_{Q}\otimes 1_{E}$ and $M\otimes V$ commute, we need to diagonalize the following matrix to solve our problem:
\begin{equation} \label{r:dephasing}
 I_{2}\otimes H_{E}+M\otimes V=
\left[
\begin{array}{cc}
 H_{E}+m_{11}V & m_{12}V   \\
 m_{12}^{\ast}V & H_{E}+m_{22}V 
\end{array}
\right].
 \end{equation} 
 The Riccati equation associated with BOM~(\ref{r:dephasing}) takes the form
 \begin{eqnarray}\label{r:mriccati}
  m_{12}XVX &+&X(H_{E}+m_{12}V) \\ \nonumber
            &-& (H_{E}+m_{22}V)X-m_{12}^{\ast}V = 0.
 \end{eqnarray}            
If $X=x1_{E}$, where $x\in\mathbb{C}$, then one can write the above equation  as
\begin{equation}
  \left(m_{12}x^2+(m_{11}-m_{22})x-m_{12}^{\ast}\right)V=0,
\end{equation} 
or equivalently as
\begin{equation}
 m_{12}x^2+(m_{11}-m_{22})x-m_{12}^{\ast}=0.
\end{equation} 
 As a result, we see that for the dephasing case the Riccati equation simplifies to the quadratic equation and therefore solution $X$ can be easily found. 
\section{Summary}\label{sec:summary}  
In this paper the problem of the exact solution of the decoherence model has been connected to the Riccati operator equation. It was shown that  obtaining the exact reduced dynamics is as problematic as resolving the Riccati equation to say the least. Furthermore, we simplified a wide class of problems described by the time dependent Hamiltonian to the time independent problems. One can easily learn from this paper that solving the time dependent Riccati~(\ref{r:Triccati}) equation is very simple.

We strongly believe that solving the model we analyzed is crucial and that it can contribute to the progress and verification of the adiabatic theorem for open quantum systems~\cite{adiab} in analogy to the contribution of the half spin particle model with Hamiltonian $H_{Q}(t,\beta)$ to the progress of the adiabatic theorem for the quantum closed systems.

\end{document}